\documentclass[twocolumn,secnumarabic,amssymb, nofootinbib, aps, prl]{revtex4-1}

\usepackage{amssymb}

\newcommand{\ba}{\begin{eqnarray}}
\newcommand{\ea}{\end{eqnarray}}
\newcommand{\be}{\begin{equation}}
\newcommand{\ee}{\end{equation}}
\newcommand{\bi}{\begin{itemize}}
\newcommand{\ei}{\end{itemize}}

\newcommand{\la}{\lambda}

\newcommand{\sa}{\sigma}

\newcommand{\cF}{{\cal F}}

\newcommand{\ra}{\rightarrow}
\newcommand{\Ra}{\Rightarrow}

\newcommand{\2}{\frac{1}{2}}

\newcommand{\Fc}{\mathcal{F}}

\begin{document}

\title{Wald's gravitational entropy for ghost-free, infinite derivative theories of Gravity}%

\author{Aindri\'u Conroy~$^{1}$, ~Anupam Mazumdar~$^{1,2}$ and Ali Teimouri~$^{1}$}
\affiliation{$^{1}$~Consortium for Fundamental Physics, Lancaster University, Lancaster, LA1 4YB, UK\\
$^{2}$ IPPP, Durham University, Durham, DH1 3LE, UK}

\begin{abstract}
In this paper, we demonstrate that the Wald's entropy for any spherically symmetric blackhole within an infinite derivative theory of gravity is determined solely by the area law. Thus, the infrared behaviour of gravity is captured by the Einstein-Hilbert term, provided that the massless graviton remains the only propagating degree of freedom in the spacetime. 
\end{abstract}

\maketitle

Einstein's general theory of relativity is a well-behaved theory of gravity in the infrared (IR), reducing to Newtonian predictions 
in the linearised limit, complete with a slowly varying source term, at large time scales and at large distances. The theory has been tested from solar-system to cosmological distances~\cite{Will}. Moreover, gravity has been tested at short distances, and there has been no departure from the 
$1/r$ fall of Newtonian potential up to $10^{-5}$~m~\cite{adel}.

One of the most intriguing properties of general relativity is that the gravitational entropy of any gravitationally bound system, as in the case of a blackhole, follows an 
area law, depicting gravity as a hologram~\cite{'tHooft:1993gx,Suss}. This has been corroborated by the Bekenstein-Hawking entropy of a blackhole~\cite{Bekenstein,Hawking}, as well as Wald's 
interpretation of gravitational entropy~\cite{Wald}. The entropy of a blackhole has been the cornerstone
of many advancements in theoretical physics, for instance, in the context of AdS (anti-de Sitter) and CFT (conformal field theory) 
correspondence~\cite{Maldacena}.

It is therefore curious to ask a question - what happens to the area law of a gravitational entropy if gravity itself gets modified in the ultraviolet (UV)?
Could there be a way to predict the form of higher order corrections in the gravitational sector from the well-known result of gravity 
being holographic? In some sense, one may ask - what kind of corrections in the (UV) in the gravitational sector would one require to maintain the holographic nature 
of gravity?

Note that the simplicity of general relativity  also leads to problems in the 
ultraviolet (UV). At short distances and at small time scales, the Ricci curvature blows up and so
too, the other observables. The theory admits well-known spacetime singularities. For instance the blackhole singularity, which is an incarnation of the Newtonian 
potential, i.e. the potential blows up close to any point source. Irrespective of the mass of the source term, the theory admits a singular solution known as 
Schwarzschild's metric within a static limit. On the other hand the theory also admits a cosmological singularity appearing at small time scales, which can be 
seen even in a homogeneous and an isotropic background solution, such as the Friedman-Lema\'itre-Robertson-Walker metric. 

The quantum corrections to the Einstein's gravity may emerge even before the $4$-dimensional Planck scale, $M_p=\sqrt{1/(8\pi G)}$.
The most general higher order action for gravity, which is generally covariant can be written (in $4$-dimensions) as follows:
\begin{eqnarray}\label{action0}
S^{tot}&=& S^{EH}+S^{UV}\nonumber \\
S^{EH} &=&\frac{1}{16\pi G}\int d^4x\ \sqrt{-g}R \nonumber \\
S^{UV}&=&\int d^4x\sqrt{-g}\left[(R^{\mu_1\nu_1\lambda_1\sigma_1}{\cal O}^{\mu_2,\nu_2\lambda_2\sigma_2}_{\mu_1\nu_1\lambda_1\sigma_1}
R_{\mu_2,\nu_2\lambda_2\sigma_2})+\cdots \right]\,,\nonumber \\
\end{eqnarray}
where the operator, ${\cal O}^{\mu_2,\nu_2\lambda_2\sigma_2}_{\mu_1\nu_1\lambda_1\sigma_1}$, contains covariant operators, such as the D'Alembertian operator $\Box=g^{\mu\nu}\nabla_{\mu}\nabla_{\nu}$.
Other contributions that are of higher order in curvature, such as cubic in curvature, quadratic in curvature, and so on and so forth are permitted by the general 
diffeomorphism.

Every operator $\Box$ comes with a scale, 
$M$, which could potentially lie anywhere between $ (10 {\rm \mu m})^{-1}\sim 100~{\rm meV}\leq M\leq 10^{19}$~GeV. In the context of string theory,
the scale $M$ could be the Kaluza-Klein scale or the compactification scale in $4$-dimensions.

Even if we restrict ourselves to the lowest order, say quadratic in curvature, there are infinitely many covariant derivatives
around Minkowski space~\cite{BGKM,BMS}.  These corrections are expected to arise very naturally in string 
field theory~\cite{SFT,SFT1}, where it is analogous to having all orders of $\alpha' $ corrections.

The aim of this letter is to compute the Wald's gravitational entropy for the above infinite higher derivative action for a static, spherically symmetric background in $4$-dimensions. In particular, this letter will establish a very intriguing link between the {\it  propagating degree of freedom} for the graviton and the gravitational entropy. The upshot is as follows: 


{As long as a higher derivative theory of gravity does not introduce any extra propagating degree of freedom, and as long as 
the IR limit of such a theory yields Einstein-Hilbert action, the contribution to the Wald's entropy due to the higher derivative corrections must vanish, yielding the famous area law of gravitational entropy, thus preserving the holographic nature of gravity. As a consequence, the gravitational entropy of 
a blackhole  for a UV  modified gravity  such as in Eq.~(\ref{action0}) will still be given by the area law.  }


We begin by noting that the above action Eq.~(\ref{action0}) can be simplified a great deal. The differential operator acting on the 
right Riemann tensor yields terms which can be integrated by parts. Couple this with the Bianchi identities and the symmetry properties of the Riemann tensor, it has been shown that the above action can be recast as ~\cite{BGKM}:
\begin{eqnarray}\label{action-main}
S^{tot} =\frac{1}{16\pi G}\int d^4x \sqrt{-g} \left[ R+\alpha \left(R \Fc_1(\Box_M)R\right.\right. \nonumber \\
\left. \left. +R_{\mu\nu}\Fc_2(\Box_M)R^{\mu\nu} + R_{\mu\nu\la\sa} \Fc_{3}(\Box_M)R^{\mu\nu\la\sa}\right) \right]\ ,
\end{eqnarray}
where $\alpha$ has inverse of mass squared dimension,and  we have defined $\Box_M\equiv \Box/M^2$ for convenience. The $\Fc$'s are the 
three unknown analytic functions given by:
\begin{equation}
\Fc_{i}(\Box_M)=\sum_{n=0}^{\infty}f_{i_n}(\Box_M)^{n}\,,
\end{equation}
where $f_{i_n}$ are appropriate constants.
The question we are keen to explore is as follows: Is there any deep connection between $\Fc$'s and the gravitational entropy? In order to address this, 
let us now consider a simple static, homogeneous and isotropic metric of the type
\begin{equation}\label{metric}
ds^2= -f(r) dt^2 + f(r)^{-1}dr^2+r^2d\Omega^2\,,
\end{equation}
where $\Omega$ denotes the angular co-ordinates. This metric has asymptotic behaviour in all three cases, i.e. Minkowski, de Sitter and anti-de Sitter.
The gravitational entropy for any such metric can be defined in terms of the Wald's entropy~\cite{Wald}, see also ~\cite{Jacobson:1993vj}. The definition of Wald's entropy follows the Bekenstein-Hawking's  
area law of a blackhole and the first law of blackhole thermodynamics, which has a clear geometric interpretation through its 
identification with the Noether charge for spacetime diffeomorphisms. The gravitational entropy can be recast as a closed integral over a cross 
section of the horizon for the metric given by Eq.~(\ref{metric}). 

For a spherically symmetric blackhole solution in $4$-dimensions, the Wald's entropy can be written as~\cite{Wald}:
\begin{equation}
S_{W}=-2\pi\oint\left(\frac{\delta{\cal L}}{\delta R_{abcd}}\right)^{(0)}\hat{\epsilon}_{ab}\hat{\epsilon}_{cd}q(r)d\Omega_{}^{2}
\end{equation}
where ${\cal L}$ is the Lagrangian, \(\hat{\epsilon}_{ab}\) is the binormal vector to the surface, where the indices $\{a,b,c,d\}\in\{r,t\}$, and
 \(q(r)d\Omega^{2}=r^{2}(d\theta^{2}+\sin^{2}\theta
d\phi^{2})\). The superscript "(0)" indicates that the
functional derivative is determined on the background
and the factor of $4$ arising due to the antisymmetric
properties of the Riemann tensor and the binormal vectors. We can then write the Wald's entropy as:
\begin{equation}
S_{W}=-8\pi\oint\left(\frac{\delta{\cal L}}{\delta R_{rtrt}}\right)^{(0)}q(r)d\Omega_{}^{2}
\end{equation}
In general, one can construct two normal directions along \(r\) and \(t\) with
\(\oint\equiv\oint_{r=r_{H},\;
t=\mbox{const}}\).
Moreover, the area of the horizon is defined to be  
\begin{equation}
\text{Area}=\oint q(r)d\Omega^{2}
\end{equation}
The Wald's entropy corresponding to  Eq.~(\ref{action-main}) can be computed by  calculating the functional
derivatives of each and every term in Eq.~(\ref{action-main}), resulting in two distinct contributions to the entropy:
\begin{equation}
S_W= S_{W}^{EH}+S_{W}^{UV}\,,
\end{equation}
given by:
\begin{eqnarray}\label{entropy-1}
\displaystyle
S_{W}=\frac{1}{4 G}\oint \left[1+\alpha \left\{2\mathcal{F}_{1}(\Box_M)R \right.\right.\nonumber \\
\left. \left. -\mathcal{F}_{2}(\Box_M) \times\left(g^{rr}R^{tt}+g^{tt}R^{rr}\right)\right. \right.\nonumber\\
\left. \left.-4\mathcal{F}_{3}(\Box_M)R^{rtrt} \right\}\right]q(r)d\Omega^{2}\,. 
\end{eqnarray}
For the metric given by Eq.~(\ref{metric}), one can see that $g_{tt}g_{rr}=-1$,  \(g_{tt}=-g^{rr}\), \(g_{rr}=-g^{tt}\). Subsequently, $(g^{rr}R^{tt}+g^{tt}R^{rr}) =-g_{ab}R^{ab}=-R$, and similarly $-2R^{rtrt}=2g_{tt}g_{rr}R^{rtrt}=g^{ab}g^{cd}R_{dacb} =R$.
With the help of these identities, we can further simplify the above expression:
 \begin{eqnarray}\label{entropy-2}
S_{W}=\frac{\mbox{Area}}{4G}\left[1+\alpha\left\{2\Fc_1(\Box_M)+\Fc_2(\Box_M)\
+2\Fc_3(\Box_M)\right\}R\right]\,.\nonumber \\
\end{eqnarray}
Interestingly, at large distances from any source term, such as in the case of IR, the action Eq.~(\ref{action-main}) is dominated by the 
Einstein-Hilbert term. It is a well known result that for the Einstein-Hilbert action, the Wald's entropy is given by
$S_{W}={\rm Area}/{4G}$. The UV part of the gravitational entropy contains a very interesting combination of $\cF$'s, which will play a crucial 
role in understanding the UV aspects of gravity and its entropy. The profound question arises - could we constrain the nature of $\cF$'s to 
some fundamental aspects of how gravity should be modified in the UV?

Let us also note,  that this modified action, Eq.~(\ref{action-main}), will inevitably modify the graviton propagator. If ${\cF}$'s contain infinite
derivatives, it would inevitably modify the graviton propagator in the UV. It is well-known that higher derivative theories have {\it ghosts} at tree-level,
for instance, the $4^{th}$ derivative gravity of Stelle's~\cite{Stelle} contains a massive ghost. Therefore, it is paramount to understand the nature of the graviton
propagator and its connection to Wald's entropy.

 The exact form of the propagator for the above action, Eq.~(\ref{action-main}),  was derived in Refs.~\cite{BGKM,Biswas:2013kla}. In principle the propagator can
be recast in terms of the spin projection operators~\cite{Van}, such as the tensor $P^{2}$ and the scalar operator
$P_{s}^0$ in the momentum space~\cite{BGKM,Biswas:2013kla}.
\be\label{prop}
\Pi(k^2)\sim {P^2\over a(k^2)k^2}+{P_{s}^0\over (a(k^2)-3c(k^2))k^2}\,,
\ee
where $a(k^2),~c(k^2)$ can be written in terms of the original $\Fc$'s contained within the modified action, as~\cite{BGKM,Biswas:2013kla}:
\begin{eqnarray}\label{a}
a(\Box_M) =   1-\frac{1}{2} {\cal F}_2 (\Box_M)\Box_M-2 {\cal F}_{3} (\Box_M) \Box_M \\
\label{c}
c(\Box_M) = 1+ 2 {\cal F}_1(\Box_M) \Box_M+ \2 {\cal F}_2 (\Box_M) \Box_M 
\end{eqnarray}
Further, since we wish to recover general relativity in the IR, i.e. $\Box\rightarrow 0,~k^2\rightarrow 0$, we must have
\be
a(0)=c(0) =1\ ,
\label{GRlimit}
\ee
 corresponding to the {\it massless} graviton propagator for the Einstein-Hilbert action. Now assuming that $a(\Box_M)=c(\Box_M)$, such that
 we can take a continuous limit from ${\rm UV}\rightarrow {\rm IR} $, we have 
\ba\label{GR-prop}
\lim_{k^2\ra 0}\Pi\sim\frac{1}{a(k^2)}\left[\frac{P^2}{k^2}-\frac{P_{s}^0}{2 k^2}\right]
\rightarrow \left[\frac{P^2}{k^2}-\frac{P_{s}^0}{2 k^2}\right]\,.
\ea
From the above Eqs.~(\ref{prop},\ref{a},\ref{c},\ref{GRlimit},\ref{GR-prop}), there are some crucial observations to make:

{\bf  Ghost free condition}: Note that UV modifications of gravity should be such that the action must have a smooth IR limit. It follows that Eq.(\ref{GRlimit}, \ref{GR-prop}) must be satisfied, i.e. the $k^2=0$ pole just describes the physical graviton state.  This also
implies that the action Eq.~(\ref{action-main}) maintains causality and the ghost free condition as long as there is no new pole introduced by the analytic function $a(\Box_M)$.
The fact that the theory must be ghost-free boils down to simply requiring that $a(\Box_M)$ is an {\it entire function}, and $a(\Box_M)-3c(\Box_M)$ has 
at most a single zero, the corresponding residue at the pole would necessarily have the correct sign, since the {\it entire function} does not have any poles in the complex plane, containing only an essential singularity at the boundary~\cite{BGKM}.

Essentially, the above Eq.(\ref{GRlimit}) means that $a(\Box_M)$ and $c(\Box_M)$ are non-singular analytic functions at $k^2=0$ and therefore cannot
contain non-local inverse derivative operators.

{\bf Constraints}: Since, we do not wish to  introduce any new extra degrees of freedom other than the massless graviton  throughout the IR to the UV, we require a constraint relationship between the $\Fc$'s from Eq.~(\ref{a},\ref{c}):
\begin{eqnarray}\label{Entrop-cond}
&&a(\Box_M) = c(\Box_M) \nonumber\\
&&\Ra 2\cF_1(\Box_M)+\cF_2(\Box_M)+2\cF_3(\Box_M)=0\,.
\end{eqnarray}
%



The above conclusions have intriguing consequences for the gravitational entropy - reducing Eq.~(\ref{Entrop-cond}) to the Wald's entropy, see Eq.~(\ref{entropy-2}), 
of a spherically symmetric blackhole, as is the case for the standard result of  $S^{EH}$:
\begin{equation}\label{main-cond}
S_{W}=S^{EH}_{W}=\frac{Area}{4G}
\end{equation}
This is an important result, the holographic nature of gravity remains preserved in spite of 
the non-trivial modifications of gravity with  infinite derivatives in the UV. The UV contribution to the gravitational entropy is simply $S^{UV}_{W}=0$, 
for a metric given by Eq.~(\ref{metric}). This result is a shear consequence of the graviton being {\it massless} and 
the requirement of not introducing of any new propagating degrees of freedom for the graviton.

This leaves us with a profound question, why is the gravitational entropy $S^{UV}_{W}=0$? As such our constraint Eq.~(\ref{Entrop-cond}) is very generic and, other than the massless nature of the graviton, 
does not shed any light on the nature of gravity in the UV. Apriori, we do not know whether
the gravitational interaction in the UV becomes weak or strong. However, the form of $\cF$'s do tell us of some interesting aspects of gravity in the UV - namely, the 
gravitational interaction becomes non-local~\cite{BMS,BGKM} and helps us to understand the quantum behaviour of gravity at higher loops, i.e. above $1$-loop
there are indications that the theory is convergent~\cite{Tomboulis,Siegel2,Modesto,Talaganis:2014ida}; explicit computations have been performed in a toy model up to 
$2$-loops~\cite{Talaganis:2014ida}. Further note that $S^{UV}_{W}=0$ will have a very interesting consequence for the the third law of thermodynamics in the context of gravity,
which might hint towards the {\it absolute ground state of gravity}, when the condition,  Eq.~(\ref{Entrop-cond}), is imposed for the action $S^{UV}$.

Gravity, being a gauge theory, contains all its interactions within the kinetic term.
If the graviton propagator is modified by Eq.~(\ref{GR-prop}), the vertex factor for any graviton-graviton interaction will also be enhanced by a factor $a(k^2)$. 
One such study has been performed in the context of singularity free gravity~\cite{Talaganis:2014ida}, where the form of $a(k^2)\sim e^{k^2/M^2}$ has been suggested, motivated by 
string field theory, where the vertex operator get similarly exponentially enhanced. In this particular case, it has been shown that the blackhole singularity for a spherically symmetric metric disappears in the linearised limit, therefore
ameliorating the UV nature of gravity~\cite{BGKM}~\footnote{The full non-linear equations of motion are extremely difficult to tackle, see~\cite{Conroy}. So far only a cosmological
solution has been constructed from the full equations of motion~\cite{BMS}.}.

At this point, however, it seems Eq.~(\ref{main-cond}) is a very generic prediction for such an infinite derivative theory of gravity, irrespective of the actual form of  $a(\Box_M)$, 
as long as $a(\Box_M)$ does not contain any additional poles.

So far, our analysis is very generic and applicable to the full  action Eq.~(\ref{action-main}). We may gain further insight into the Newtonian potentials of the metric
by assessing the linearised metric, such that it becomes an asymptotically flat spacetime. Let us assume that the $(t,~r)$ component of the original spherically symmetric 
metric, Eq.~(\ref{metric}), takes the form:
\begin{equation}\label{lin-met}
ds^{2}=-(1+2\Phi(r) )dt^{2}+(1-2\Psi (r)) dr^{2}\,,
\end{equation}
where $2\Phi(r),~2\Psi(r) \ll 1$. In fact, $\Phi$ and $\Psi$ are the two Newtonian potentials.
One can then ask: What should be the Wald's entropy in the linearised limit of the action given by Eq.~(\ref{action-main})?
For the above metric, Eq.~(\ref{lin-met}), we can evaluate the Wald's entropy, simplifying the expression by 
considering a static solution. The gravitational entropy  is then given, at the linearised order, by
\begin{eqnarray}
S_{W}=\frac{Area}{4 G}\left\{1+ 2\Psi-2\Phi \right.\nonumber \\
\left.
+ \alpha\left[2{\cF}_1(\Box_M)+{\cF}_2(\Box_M)+2{\cF}_3(\Box_M)\right](-2\Phi^{\prime\prime})\right\}\,.
\end{eqnarray}
where $^\prime$ denotes the derivative with respect to $r$. Note that when
$\Psi  = \Phi $ and  $2{\cF}_1+{\cF}_2+2{\cF}_3 = 0$, 
for any source term within the linearised limit, the gravitational entropy 
duly reduces to that of the $S^{EH}_{W}$. The conditions are exactly the same as that of our complete analysis.
Indeed, it would be interesting to seek scenarios when the area law of a blackhole might incur modifications. 
One might imagine departing from the assumption of spherical symmetry, in which case it is possible to realise $\Phi\neq \Psi$. 

However, the other possibility, when $a(\Box_M)\neq c(\Box_M)$, is more interesting. This condition would immediately imply that there are additional poles in the graviton propagator, other than the massless graviton, 
see Eqs.~(\ref{prop},~\ref{GR-prop}). For instance, ${\cal L}\sim f(R)$ gravity, which is very popular due to its simplicity contains an extra scalar degree of freedom other than the 
massless graviton, see~\cite{Biswas:2013kla}. Any UV modification with $f(R)$ gravity would therefore contribute to the gravitational entropy besides $S_{W}^{EH}$. However, 
such class of gravity does not ameliorate the UV aspects of gravity at all~\cite{BGKM}. Similarly, the conformal invariant gravity, ${\cal L}=R-\alpha C^2$ contains massive 
spin-2 degree of freedom other than the massless graviton~\cite{Biswas:2013kla,BGKM}. Moreover, this massive spin $2$ degree of freedom comes with a wrong 
sign in the graviton propagator, thus revealing a massive ghost. Both these examples are a subset of the above action Eq.~(\ref{action-main}), and suggest that one of $\cF$'s should be zero.
Of course, a consequence of such a vanishing function $\cF$ is that the action would be incomplete from the UV point of view.

Before we conclude, let us briefly bring the reader's attention to this final intriguing point. The condition $a(\Box_M)\neq c(\Box_M)$ seems to have some relevance for cosmology. Unlike the blackhole case the cosmological singularity cannot be avoided by assuming
$a(\Box_M)= c(\Box_M)$ as shown in~\cite{BMS,BKM,others}. One requires additional degrees of freedom other than the massless graviton, which remains a tantalising issue - leading one to ask why the respective natures of these two singularities are so different? And why the fundamental nature of the graviton has to deviate to understand these two problems ?

In conclusion, we have found a very intriguing result for a class of ghost-free, infinite derivative theory of gravity - the gravitational entropy for a spherically 
symmetric metric is solely given by the Einstein-Hilbert action. The area law of gravitational entropy is the main contribution arising from the IR aspect of gravity, while the 
UV contribution ( from an action up to quadratic in curvature ) of the gravitational entropy vanishes exactly.  This happens due to an interesting connection between the 
propagating degrees of freedom for the graviton - if the massless graviton remains the only propagating degree of freedom in the spacetime then there will be no other contribution to the gravitational entropy other than the Einstein-Hilbert term's contribution. In generality, at least in the spherically symmetric case, gravity remains holographic! Our result, $S^{UV}_{W}=0$, will have some deep consequences for the third law of thermodynamics in the gravitational system, which we shall explore in the future.

*{Acknowledgement} AC is funded by STFC grant no ST/K50208X/1, and AM is supported by the STFC grant ST/J000418/1. AM acknowledges the kind hospitality from  IPPP, Durham, 
during the course of this work.

\end{document}